\documentclass[reprint,superscriptaddress,amssymb,aps, pra, twocolumn,graphicx]{revtex4-1}

\usepackage{graphicx}
\usepackage{dcolumn}
\usepackage{bm}
\usepackage{hyperref}

\usepackage{amsmath} 
\usepackage{mathrsfs}
\usepackage[cmintegrals]{newtxmath}
\usepackage[mathlines]{lineno}

\begin{document}


\title{Experimental demonstration of phase-matching and Sagnac effect in a millimeter-scale wedged resonator gyroscope}

\author{Xuan Mao}
\email{These authors contributed equally to this work}
\affiliation{Department of Physics, State Key Laboratory of Low-Dimensional Quantum Physics, Tsinghua University, Beijing 100084, China}

\author{Hong Yang}
\email{These authors contributed equally to this work}
\affiliation{Department of Physics, State Key Laboratory of Low-Dimensional Quantum Physics, Tsinghua University, Beijing 100084, China}

\author{Dan Long}
\affiliation{Department of Physics, State Key Laboratory of Low-Dimensional Quantum Physics, Tsinghua University, Beijing 100084, China}

\author{Min Wang}
\email{wangmin@baqis.ac.cn}
\affiliation{Beijing Academy of Quantum Information Sciences, Beijing 100193, China}

\author{Peng-Yu Wen}
\author{Yun-Qi Hu}
\author{Bo-Yang Wang}
\affiliation{Department of Physics, State Key Laboratory of Low-Dimensional Quantum Physics, Tsinghua University, Beijing 100084, China}

\author{Gui-Qin Li}
\author{Jian-Cun Gao}
\affiliation{Department of Physics, State Key Laboratory of Low-Dimensional Quantum Physics, Tsinghua University, Beijing 100084, China}
\affiliation{Frontier Science Center for Quantum Information, Beijing 100084, China}

\author{Gui-Lu Long}
\email{gllong@tsinghua.edu.cn}
\affiliation{Department of Physics, State Key Laboratory of Low-Dimensional Quantum Physics, Tsinghua University, Beijing 100084, China}
\affiliation{Beijing Academy of Quantum Information Sciences, Beijing 100193, China}
\affiliation{Frontier Science Center for Quantum Information, Beijing 100084, China}
\affiliation{Beijing National Research Center for Information Science and Technology, Beijing 100084, China}
\affiliation{School of Information, Tsinghua University, Beijing 100084, China}

\date{\today}

\begin{abstract}

The highly efficient coupling of light from conventional optical components to optical mode volumes lies in the heart of chip-based micro-devices, which is determined by the phase-matching between propagation constants of fiber taper and the whispering-gallery-mode (WGM) of the resonator. Optical gyroscopes, typically realized as fiber-optic gyroscopes and ring-laser gyroscopes, have been the mainstay in diverse applications such as positioning and inertial sensing. Here, the phase-matching is theoretically analyzed and experimentally verified. We observe Sagnac effect in a millimeter-scale wedged resonator gyroscope which has attracted considerable attention and been rapidly promoted in recent years. We demonstrate a bidirectional pump and probe scheme, which directly measures the frequency beat caused by the Sagnac effect. We establish the linear response between the detected beat frequency and the rotation velocity. The clockwise and counterclockwise rotation can also be distinguished according to the value of the frequency beat. The experimental results verify the feasibility of developing gyroscope in WGM resonator system and pave the way for future development.

\end{abstract}


\maketitle


\section{INTRODUCTION \label{introduction}}

Chip-based devices are a mainstream of development in recent years, such as photon source \cite{wang2019chemo}, optical sensing \cite{qin2021experimental}, and quantum key distribution \cite{semenenko2020chip, wang2020quantum}. Gyroscopes for rotation sensing with high sensitivity and precision have been studied in various systems ranging from micro-optical-electro-mechanical system \cite{chang2008integrated, alper2005single}, cold atom system \cite{stringari2001superfluid, dutta2016continuous}, optomechanical system \cite{davuluri2017gyroscope, mao2020enhanced}, photon and matter-wave interferometers \cite{haine2016mean, scully1993quantum}, to solid spin system \cite{kornack2005nuclear, jaskula2019cross, wood2017magnetic}. Based on the Sagnac effect \cite{post1967sagnac}, the phase difference between two counter-propagation beams in a resonator resulting from rotation, optical gyroscopes \cite{jin2018short, liang2017resonant, matsko2018fundamental, zhang2017single, wang2015resonator, tian2019rotation} have been widely used in numerous applications such as remote control \cite{jaroszewicz2011afors} and optical test for gravitation theories \cite{sokolov2015development} in recent decades. Therefore, observation of the Sagnac effect is a crucial step to perform rotation sensing and provides the feasibility of developing gyroscope in optical system.

Whispering-gallery-mode (WGM) resonators as a crucial branch of optical systems have become a promising platform for both fundamental research \cite{wang2019characterization,  jiang2015chip} and practical applications \cite{jiang2017chaos, lu2015p, zhang2017far, xu2020frequency, duggan2019optomechanically, mao2022tunable, hu2021demonstration, liang2021low} owing to the ability to enhance light-matter interaction in an ultra-small volume. Among these applications, WGM resonators play a significant role in sensing of different physical quantities such as nanoparticles \cite{chen2017exceptional, qin2021experimental, qin2019brillouin}, pressure \cite{ioppolo2007pressure}, acceleration \cite{li2018characterization}, and rotation velocity \cite{lai2020earth, wang2020petermann, li2017microresonator, an2015high}. The highly efficient coupling \cite{knight1997phase, farnesi2014long, shi2022two, wang2018optothermal} of light from conventional components to optical modes is a crucial factor in the development of chip-based devices. Besides, the fabrication of WGM resonators is compatible with the traditional semiconductor material processing, which promises the integration of optical gyroscopes based on WGM resonators. Advances in microfabrication techniques, combined with various methods of measuring Sagnac effect in optical resonators \cite{li2017microresonator, an2015high, tian2018rotation}, show potential in developing chip-based, mass-producible optical gyroscopes. As a result, a high quality (Q) factor WGM resonator becomes the key sensing element of optical gyroscopes. However, developing high precision gyroscopes puts forward a big challenge for resonator fabrication and the universal schemes for gyroscopes are needed.

In this paper, we theoretically analyze and experimentally demonstrate the phase-matching between the propagation constants of the fiber taper and the WGM in resonators which are associated with geometry dimensions. The results show that both the coupling efficiency and the Q factor of the coupling system exhibit a down parabola-like distribution as the diameter of the fiber taper increases for a given size resonator. Using the high Q wedged resonator, we propose a bidirectional pump and probe scheme to verify Sagnac effect. According to the theoretical formula of Sagnac effect, the linear correspondence between the detected frequency beat and the rotation velocity can be established in the experiments. Furthermore, the proposed resonator gyroscope scheme can distinguish the clockwise and the counterclockwise rotation directions according to the value of the frequency beat. The demonstrated resonator gyroscope configuration differs from the conventional schemes in the following aspects: (1) Different from the previous proposals, the proposed bidirectional pump and probe scheme is a universal scheme that does not rely on some specific effect of the resonator. In this experiment, the sensing unit includes a wedged resonator with a diameter of $2.5$ mm and a Q factor of about $6 \times 10^6$, which is available in conventional fabrication processing. (2) The design of the optical gyroscope based on the wedged resonator is highly integrated benefiting from the compatible fabrication technique of the wedged resonator. The experimental results verify Sagnac effect in the wedged resonator and provide the feasibility of developing gyroscopes in the WGM resonator system. The proposed configuration may inspire new technological developments in various schemes and provide a practical approach for the development of other classes of integrated sensors.

This article is organized as follows: In Sec.\ref{phase}, we demonstrate phase-matching between wedged resonator and the fiber taper. We observe the Sagnac effect in Sec.\ref{gyroscope}. Conclusion is given in Sec.\ref{conclusion}.

\section{PHASE-MATCHING BETWEEN WEDGED RESONATOR AND FIBER TAPER\label{phase}}

Consider a whispering-gallery-mode (WGM) resonator with a radius of $R_0$ and a uniform refractive index of $n_r$, which is surrounded by a background medium of index $n_0$. The WGM supported in the resonator can be described by three parameters $l$, $m$, and $n$, which represents the angular mode number,  azimuthal mode number and radial mode number respectively. In integrated optics, the propagating constant \cite{little1999analytic} of the WGM in the resonator, the wave vector in the net direction of propagation $\beta_m$, can be expressed as

\begin{align}
  \beta_m = \frac{m}{R_0}. \label{Equation 1}
\end{align}

For a fiber taper with a core radius of $a$, a core refractive index of $n_f$ and a cladding index of $n_{cl}$, the characteristic equation \cite{little1999analytic} to determine the propagation constant $\beta_f$ in the assumption of the linearly polarized fields is

\begin{align}
  k_f \frac{J_1(a k_f)}{J_0(a k_f)} = \alpha_f \frac{K_1(a \alpha_f)}{K_0(a \alpha_f)}, \label{Equation 2}
\end{align}
where
\begin{align}
  k_f      & = \sqrt{k^2 n_f^2 -\beta_f^2}, \label{Equation 3}    \\
  \alpha_f & = \sqrt{\beta_f^2 -k^2 n_{cl}^2}, \label{Equation 4}
\end{align}
where $k$ represents the wave vector, $J_0$ and $J_1$ are the Bessel functions of zero and first order, $K_0$ and $K_1$ are the modified Hankel functions of zero and first order, respectively.

In the experiments, the WGM of the resonator can be excited by a fiber taper. The intracavity power $\kappa^2$ coupled from the fiber taper into the resonator can be described by

\begin{align}
  \kappa^2 = \frac{1}{\beta_f^2} \kappa_0^2 \frac{\pi}{\gamma_t}e^{-\Delta\beta^2 / (2 \gamma_t)}, \label{Equation 5}
\end{align}
where
\begin{align}
  \Delta\beta = \beta_f - \beta_m, \gamma_f = \alpha_f \frac{K_1(a \alpha_f)}{K_0(a \alpha_f)}, \gamma_t = \frac{\gamma_f}{2 R_0}, \label{Equation 6}
\end{align}
where $\Delta\beta$ is the difference of propagation constants of the fiber mode and the WGM. $\kappa_0^2$ is associated with the fiber mode field and the WGM field while is independent with $\Delta\beta$. It can be inferred that the phase-matching between the fiber mode and the WGM in resonator plays a crucial role in obtaining high coupling efficiency and high Q factor. As a conclusion of Eq. \ref{Equation 1} and Eq. \ref{Equation 2}, there is also a matching condition between the diameters of the fiber taper and the WGM resonator.

\begin{figure}
  \centering
  \includegraphics[width=\linewidth]{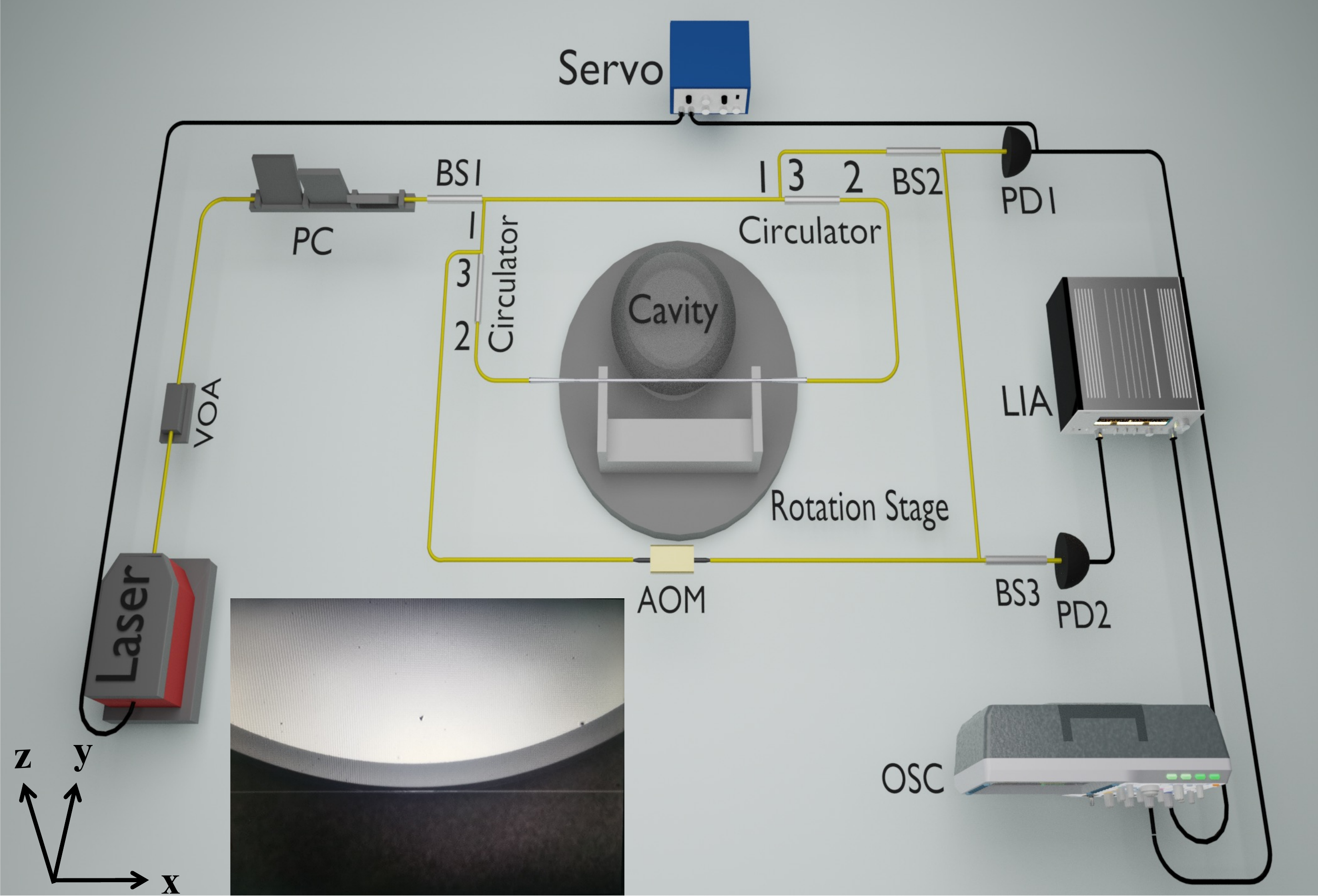}
  \caption{The optical circuit of the experiment. VOA, variable optical attenuator; PC, polarization controller; BS, beam splitter; AOM, acoustic optical modulator; PD, photodetector; OSC, oscilloscope; LIA, lock-in amplifier. The inset illustrates the top view of the coupling between the wedged resonator and the fiber taper.}
  \label{setup}
\end{figure}

The optical circuit used in the experiment is shown in Fig. \ref{setup}.
The experimental setup possess two different functions: (1) Performing Q factor measurement of the wedged resonator and the fiber taper when acoustic optical modulator (AOM) is not driving by external radio frequency. (2) Testing the Sagnac response of the resonator gyroscope when AOM is driving by external radio frequency.

To perform Q factor measurement, a tunable narrow linewidth external-cavity laser works as probe laser with emission in $1550$ nm. Before coupling into the wedged resonator, the power and the polarization of the light are manipulated by a variable optical attenuator and a polarization controller. The single mode fiber taper is used in this experiment to couple the light in and out of the wedged resonator with a diameter of $2.5$mm. In order to precisely control the relative position of the fiber taper and the wedged resonator in three dimensions, the resonator is placed on a piezoelectric stage and a microscope is mounted in front of the resonator to monitor the relative position. The output of the light at the other end of the fiber taper is detected by a photodetector. To measure the Q factor of the resonator, the frequency of the laser is scanning which is controlled by a sweep signal from function generator.

\begin{figure}
  \centering
  \includegraphics[width=0.9\linewidth]{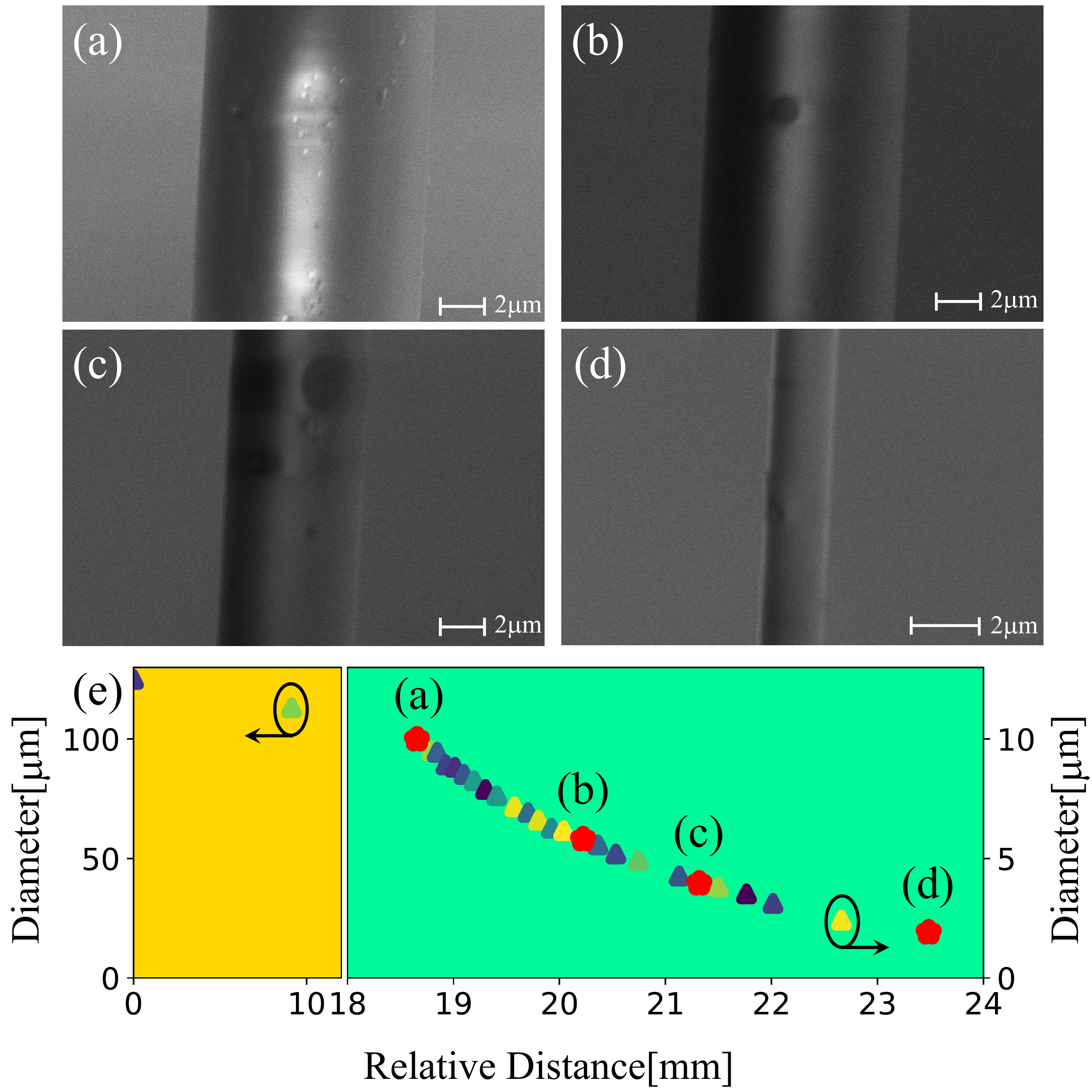}
  \caption{Characterization of the fiber taper. (a) - (d) are the side view scanning electron microscope (SEM) images of the fiber taper. (e) illustrates the diameters of the fiber taper versus relative distances to the marked point in the experiment along the x-axis shown in Fig. \ref{setup}.}
  \label{fiber}
\end{figure}

Observing the phase-matching between the fiber taper and the wedged resonator, we couple the resonator with different position along the fiber taper. The diameter distribution of the fiber taper is illustrated in Fig. \ref{fiber} (e) and every dot corresponds to a Q factor measurement of the resonator. As the relative distance along the x-axis shown in Fig. \ref{setup} increases, the diameter of the fiber taper decreases. To show the diameter change visually, we pick four dots in Fig. \ref{fiber} (e) marked by the red marker and their side view scanning electron microscope images are shown in Fig. \ref{fiber} (a) - Fig. \ref{fiber} (d). The longer the relative distance is, the thinner the diameter of the fiber taper is.

\begin{figure}
  \centering
  \includegraphics[width=\linewidth]{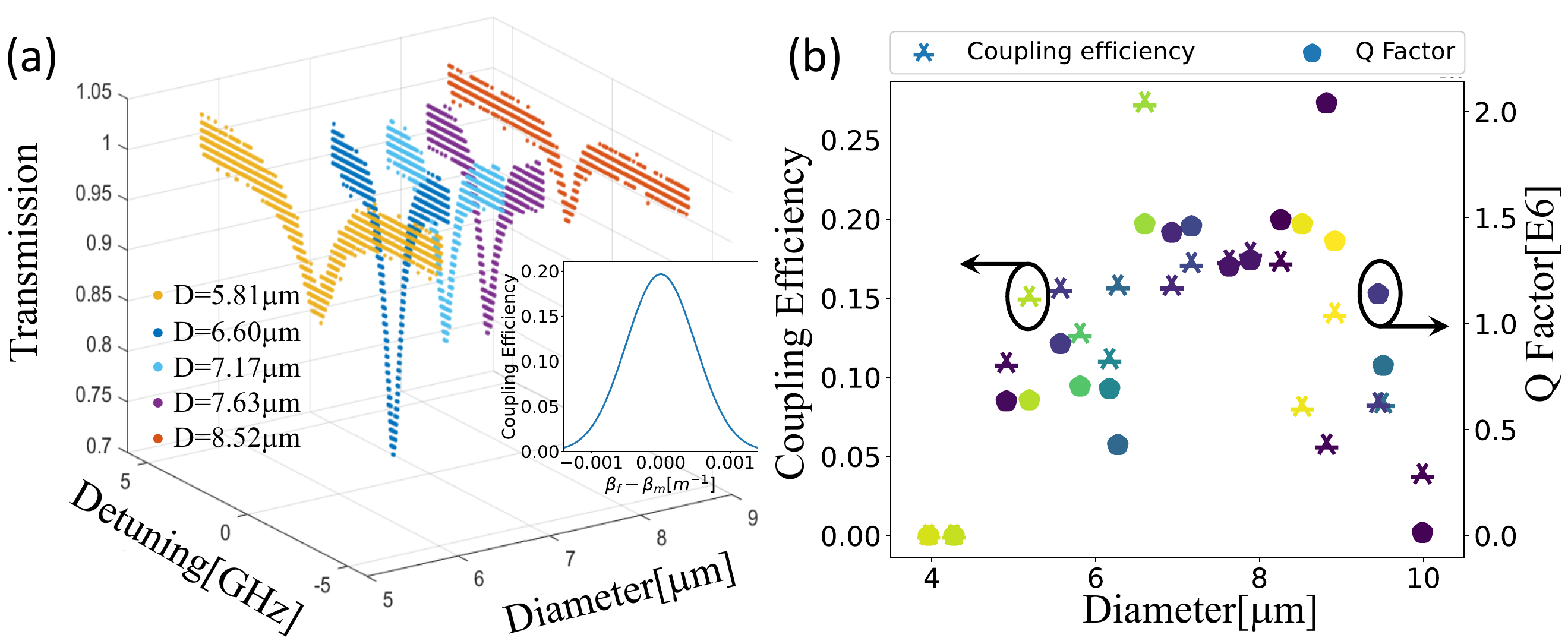}
  \caption{Experimental results. (a) Transmission spectra of the wedged resonator coupled with different positions of the fiber taper characterized by different diameters. The inset illustrates the theoretical coupling efficiency demonstrated by Eq. \ref{Equation 5} versus the difference of propagation constants of the fiber taper and the WGM with $m = 10$, $R_0 = 1.25$mm, $\gamma_t = 2.5 \times 10^-7 m^{-2}$ , and $\kappa_0 = 1$. (b) The coupling efficiency and Q factor ($@1548.92$nm) versus the diameters of the fiber taper.}
  \label{3D}
\end{figure}

To demonstrate the influence of the diameter of the fiber taper on the coupling efficiency, the resonator coupling with different positions along the fiber taper and the transmission spectra exhibit different linewidth and coupling depth. Fig. \ref{3D} (a) shows the transmission spectra behavior as the diameter of the fiber taper increase where the diameters are $5.81$ $\mu$m, $6.60$ $\mu$m, $7.17$ $\mu$m, $7.63$ $\mu$m, and $8.52$ $\mu$m, respectively. Furthermore, the coupling efficiency and the loaded Q factor of the resonator-taper coupling system for the same WGM mode ($@1548.92$nm) is shown in Fig. \ref{3D} (b) as the diameter increases. It can be concluded from Fig. \ref{3D} (a) that the coupling depth undergoes an increase followed by a decrease as the taper diameter increases, and there exists an optimal diameter for the maximum coupling efficiency, when the phase-matching condition of the propagation constants is satisfied. The experimental results are in good agreement with the prediction of Eq. \ref{Equation 5}, which is shown in the inset of Fig. \ref{3D} (a).

The diameter of the fiber taper also has significant impact on the loaded Q factor of the coupling system for the same WGM mode. Fig. \ref{3D} (b) depicts the total Q factor of the coupling system for the same WGM mode ($@1548.92$nm) as the diameter of the fiber taper increases. Similar to the behavior of the coupling efficiency, the measured Q factor of the WGM mode experiences increase first and then decrease processing. There is also an optimal taper diameter to obtain the highest loaded Q factor within the same optical mode. In our experiment, the highest Q factor of $2\times10^6$ is obtained when coupling the resonator with fiber waist of $8.62\mu$m.

\section{SAGNAC EFFECT IN WEDGED RESONATOR GYROSCOPE\label{gyroscope}}

The Sagnac effect is the basis of rotation sensing in optical resonator. In an optical resonator of radius $r$ and refractive index $n$, the Sagnac-Fizeau frequency shift \cite{malykin2000sagnac} the intra-cavity laser experienced with the angular velocity $\Omega$ can be expressed as
\begin{align}
  \omega_a        & \rightarrow \omega_a \pm \Delta_{Sagnac}, \label{equation 7}                                                                          \\
  \Delta_{Sagnac} & = \frac{n r \Omega \omega_{a}}{c} (1 - \frac{1}{n^2} - \frac{\lambda}{n} \frac{\mathrm{d} n}{\mathrm{d} \lambda}), \label{equation 8}
\end{align}
where $\omega_a$ and $\lambda$ are the frequency and the wavelength of the laser with the absence of rotation, respectively. $c$ is the speed of light in vacuum and the dispersion term $dn / d\lambda$, $\sim 1\%$ for typical material \cite{lu2017optomechanically}, is negligible. Under the condition that the rotation speed is much smaller than the light speed, the relativistic effects can be ignored.

The gyroscope sensing unit includes the high-Q wedged resonator with a diameter of $2.5$ mm and a Q factor greater than $6 \times 10^6$. The optical circuit of the experiment is demonstrated by Fig. \ref{setup}. Light from a tunable external cavity diode laser at around $1550$ nm is modulated by a variable optical attenuator and a polarization controller to adjust the power and the polarization state before being coupled into the wedged resonator with the aid of a tapered fiber. A 50/50 beam splitter (BS) divides the light into two equal parts and then couple into the same fiber taper with the opposite directions using two circulators. The opposite propagation directions in the fiber lead to the clockwise (CW) and the counterclockwise (CCW) propagation pairs in resonator. After introducing rotation into the system, the two propagating lights experience the same Sagnac frequency shift but in opposite directions. The frequency locking unit is composed of a $10/90$ BS2, a photodetector (PD1) and a servo controller. Since pumping the wedge resonators in both directions with the same frequency would cause interference problems, which are destructive to the frequency locking unit, an acoustic optical modulator (AOM) is applied on one arm before the combination of the signals. PD2 converts optical signals into electrical signals to a lock-in amplifier to monitor the beat frequency. Furthermore, applying AOM in one of the two light paths would respond differently to different rotation directions.

Combining the above analysis and Fig. \ref{setup}, one can find the different pathways from BS1 to BS3: \textcircled{1} the light from BS1 and circulator transmits through fiber taper from the right hand while does not couple into the cavity. \textcircled{2} the light from BS1 and circulator transmits through fiber taper from the right hand and couple into the cavity. \textcircled{3} the light from BS1 and the other circulator transmits through fiber taper from the left hand while does not couple into the cavity. \textcircled{4} the light from BS1 and the other circulator transmits through fiber taper from the left hand and couple into the cavity. Before BS3, the light of the pathways \textcircled{1} and \textcircled{2} experiences additional frequency shift $f_{AOM}$ due to the presence of the AOM in the circuit. In the case of CW rotation, the frequency of the above pathways are $\omega_a + f_{AOM}$, $\omega_a - \Delta_{Sagnac} + f_{AOM}$, $\omega_a$, and $\omega_a + \Delta_{Sagnac}$, respectively. Thus the beat frequency can be $\Delta_{Sagnac}$, $f_{AOM}$, $f_{AOM} - 2 \Delta_{Sagnac}$, and $f_{AOM} - \Delta_{Sagnac}$. While in the case of CCW rotation, the frequency of the above pathways are $\omega_a + f_{AOM}$, $\omega_a + \Delta_{Sagnac} + f_{AOM}$, $\omega_a$, and $\omega_a - \Delta_{Sagnac}$, respectively. Thus the beat frequency can be $\Delta_{Sagnac}$, $f_{AOM}$, $f_{AOM} + \Delta_{Sagnac}$, and $f_{AOM} + 2 \Delta_{Sagnac}$. In the experiments, the beat frequencies $\Delta_{Sagnac}$ can be obtained for different rotation velocities and directions. Thus, one can establish linear response between the detected beat frequency and the rotation velocity. In a conclusion, the resopnses of the optical circuit depend on the rotation direction and the proposed resonator gyroscope has the ability of distinguishing CW rotation and CCW rotation.

\begin{figure}
  \centering
  \includegraphics[width=\linewidth]{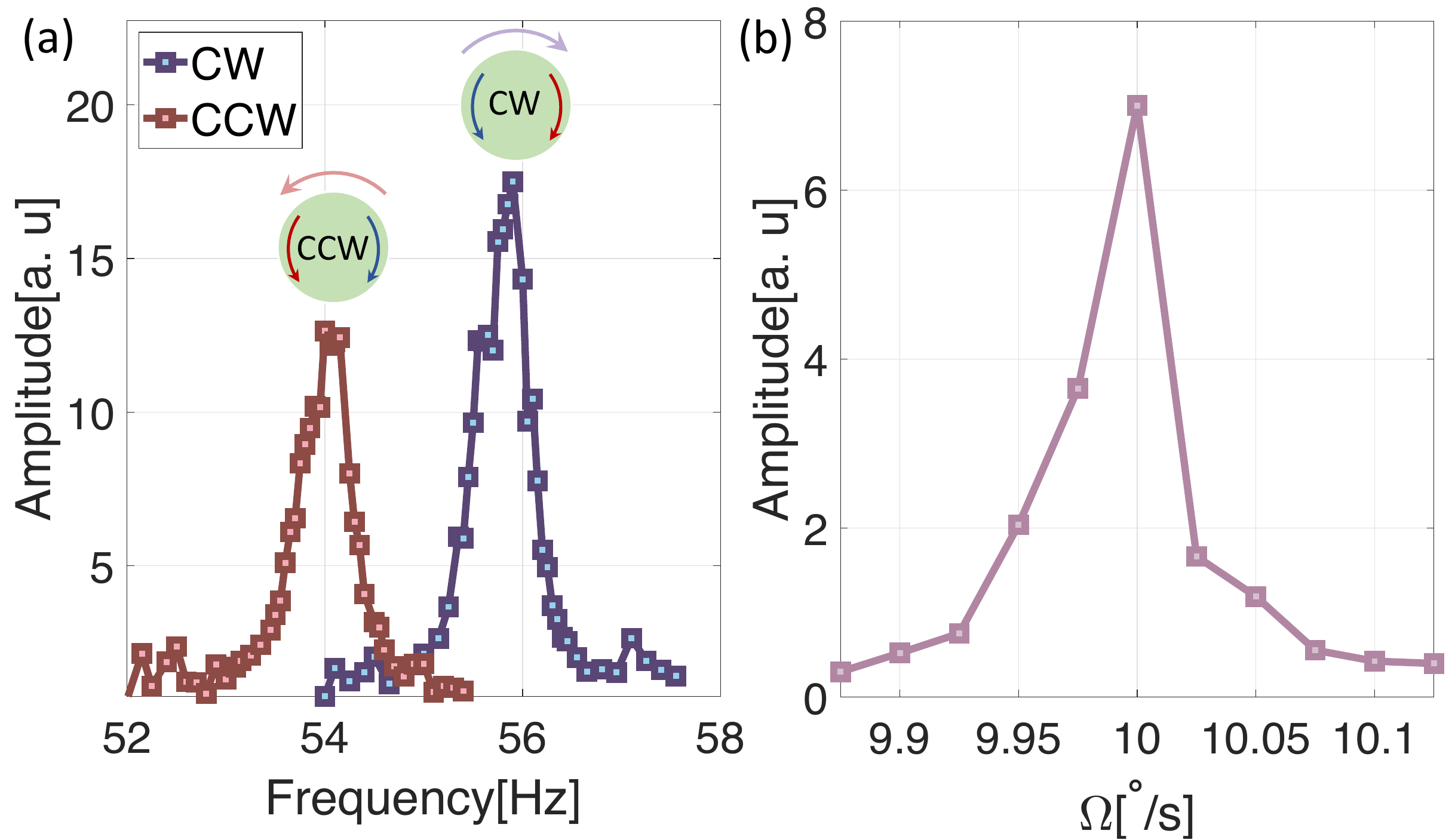}
  \caption{Experimental results. (a) The gyroscope readouts versus LIA reference frequency for different rotation directions. (b) The gyroscope readouts versus rotation velocity with fixed LIA reference frequency and the same rotation direction.}
  \label{scan}
\end{figure}

To test resonator gyroscope, a fixed rotation frequency or direction is introduced into the system. And the experimental results are demonstrated by Fig. \ref{scan} (a). The gyroscope readouts vary with LIA reference frequency and the highest amplitude is corresponding to the most appropriate LIA reference frequency. The insets in Fig. \ref{scan} (a) illustrate the intra-cavity lights experiencing the Sagnac frequency shift when the resonator rotates in the CW and CCW direction. The red (blue) arrow indicates the propagating light have red (blue) frequency shift.

On the other hand, we observe the gyroscope readouts in Fig. \ref{scan} (b) with fixed LIA reference frequency, the same rotation direction and different rotation velocity. Similar to Fig. \ref{scan} (a), the amplitude trends to increase as the rotation velocity increases before a specific velocity $\Omega_c$. When the rotation velocity exceeds $\Omega_c$, the amplitude of the gyroscope readouts begins to decrease and the faster the rotation velocity is, the smaller the amplitude is. Based on Fig. \ref{scan} (b), we can find the best matched velocity $\Omega_c$ with a specific reference frequency and rotation direction.

Combining the results of Fig. \ref{scan} (a) and Fig. \ref{scan} (b), we can establish a one-to-one correspondence between LIA reference frequency and rotation velocity. Benefiting from the relatively narrow linewidth of the LIA filter, it is more accurate for us to establish the one-to-one correspondence. The best-matched reference frequency reflects the frequency beat due to Sagnac effect in the resonator.

\begin{figure}
  \centering
  \includegraphics[width=\linewidth]{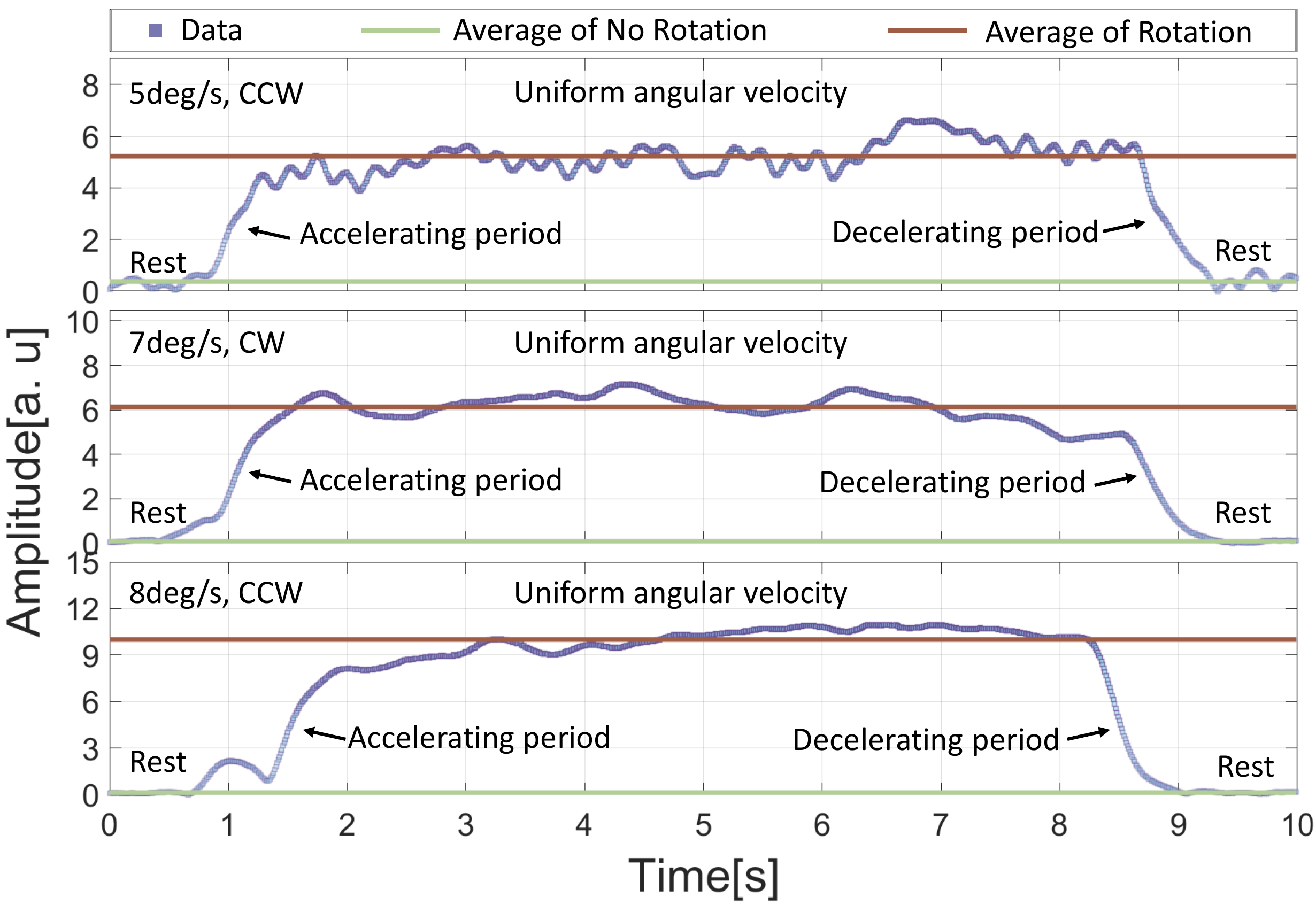}
  \caption{Experimental results. The gyroscope readouts when the wedged resonator experiences rest-accelerating-rotate uniformly-decelerating-rest five states for different rotation velocities and different directions. From top to bottom, the rotation modes are $5$ deg/s CCW, $7$ deg/s CW and $8$ deg/s CCW, respectively.}
  \label{rotations}
\end{figure}

We investigate the gyroscope readouts for different rotation velocities and directions, the experimental results are shown in Fig. \ref{rotations}. The resonator experiences rest-accelerating-uniform angular velocity-decelerating-rest five states in every experiment. From top to bottom, the rotation modes are $5$ deg/s CCW, $7$ deg/s CW and $8$ deg/s CCW, respectively. Using the one-to-one correspondence as mentioned above, the reference frequency is set to the best-matched value. As the frequency beat is proportional to the rotation frequency as indicated in Eq. \ref{equation 8}, the amplitude variation behaviors are similar to the change of rotation velocity as shown in Fig. \ref{rotations}. The purple boxes in the figure are the data points. The red lines represent the average value of the output signal of the gyroscope in the presence of the uniform angular velocity and the green lines represent the average value of the output signal when the gyroscope is at rest. According to the amplitude value, the optical resonator gyroscope can distinguish between static and rotating.

\begin{figure}
  \centering
  \includegraphics[width=\linewidth]{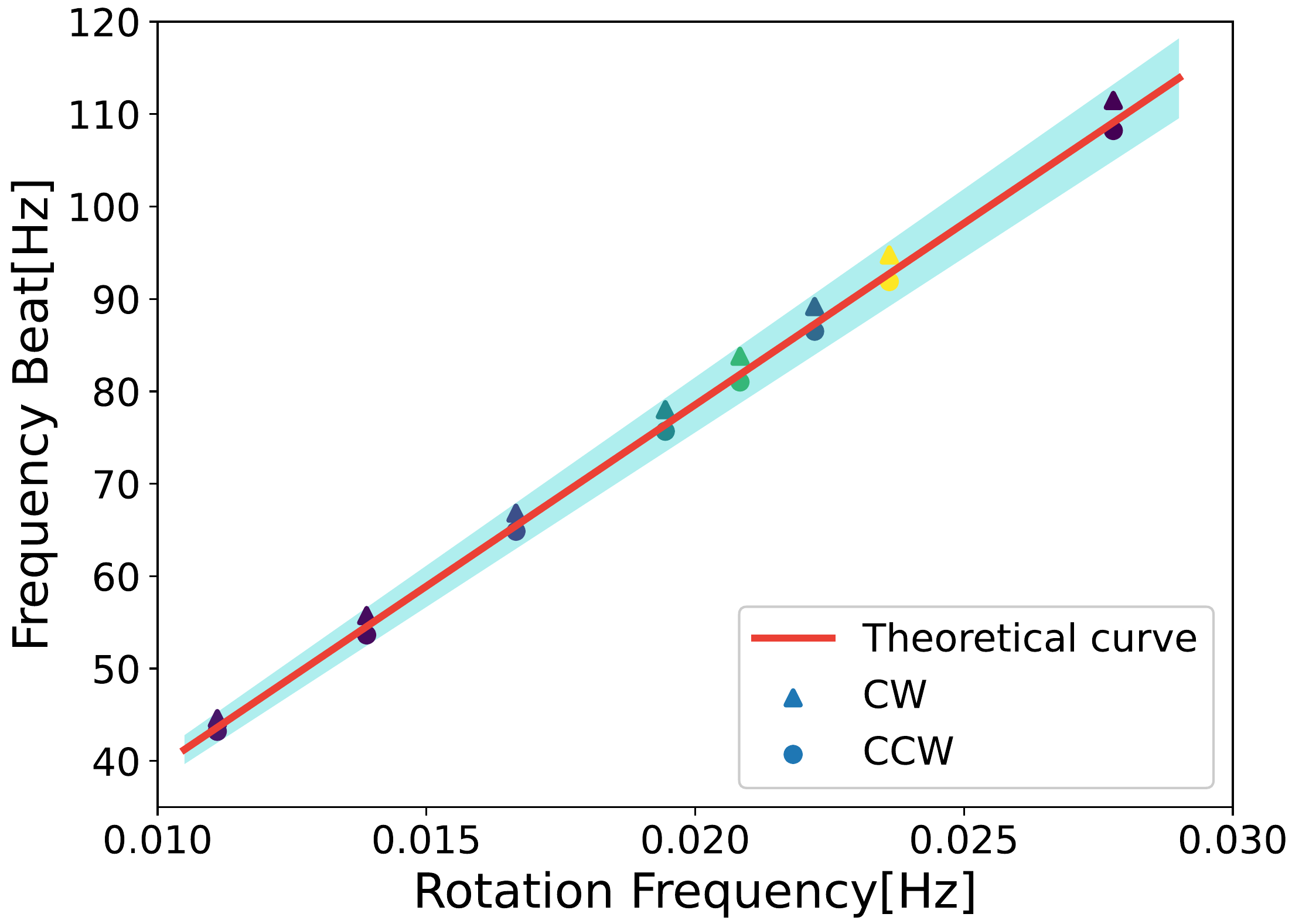}
  \caption{The detected frequency beat versus rotation frequency. CW and CCW rotation directions are marked by triangles and circles, respectively. The solid line denotes the theoretical curve presented by Eq. \ref{equation 8}.}
  \label{sagnac}
\end{figure}

In order to accurately verify the Sagnac effect in the optical cavity, we detect the frequency beats at different speeds and the results are shown in Fig. \ref{sagnac}. The detected frequency beats of CW and CCW rotation directions are marked by triangles and circles, respectively. The solid line shows the theoretical curve dominated by Eq. \ref{equation 8}. The parameter values used here are $n=1.46$, $r=1.25$ mm, $c=3 \times 10^8$ m/s, $\lambda = 1550$ nm. The band in Fig. \ref{sagnac} corresponds that the effective refractive index of silicon dioxide ranges from $1.44$ to $1.48$ as it is affected by several environmental parameters such as temperature and humidity. The results show that the frequency beats is proportional to the rotation frequency no matter which direction the gyroscope rotates in, which verifies that the frequency beat we observed is indeed caused by Sagnac effect. The experimental results are in good agreement with the theoretical curve benefiting from the narrow linewidth the filter of the LIA owns. 

\section{CONCLUSION AND OUTLOOK\label{conclusion}}

In this paper, the phase-matching between the propagation constants associated with geometry dimensions of the fiber taper and the WGM in resonator is theoretically analyzed and experimental verified. We have demonstrated a bidirectional pump and probe scheme to directly measure the frequency beat caused by the rotation of the counter-propagating light pairs in resonator. Benefiting from the stability of laser frequency locking and the advantages of narrow linewidth of LIA, we establish the linear correspondence between the detected beat frequency and the rotation velocity. On the other hand, due to introducing the asymmetry operation of the two lightpathes, the CW rotation and CCW rotation can be distinguished according to the value of the frequency beat. Furthermore, we implement experiments with some specific rotation velocity and two different rotation directions and the results show that the frequency beat is proportional to the rotation frequency no matter which direction the gyroscope rotates in, which verifies that the frequency beat we observed is indeed caused by Sagnac effect. The frequency beats we detected have good agreement with the theoretical curve. It is believed that the bigger wedged resonator and the higher quality factor promise lower detectable rotation rates. In the past decades, lots of technique methods emerge to reinforce the performance of gyroscopes such as exceptional point \cite{lai2019observation, wang2020petermann, mao2020enhanced, grant2020enhanced} and exceptional surfaces \cite{qin2021experimental, li2021exceptional, yang2021scalable}, parity-time symmetry \cite{peng2014parity, chang2014parity} and anti-parity-time symmetry \cite{zhang2020breaking, qin2021sensing}, and nonlinear enhancement \cite{zhang2020breaking, silver2021nonlinear, rodriguez2020enhancing}. Combining these methods, the precision and the sensitivity of the resonator gyroscope can be further reinforced. The experimental results demonstrated in the main text verified the feasibility of developing gyroscope in whispering-gallery-mode resonator system. The proposed configuration may inspire new technological developments in various schemes in which measuring low rotation rates via ultra-compact systems is highly attractive.

\begin{acknowledgments}

This work is supported by the National Natural Science Foundation of China (61727801, and 62131002), National Key Research and Development Program of China (2017YFA0303700), Special Project for Research and Development in Key areas of Guangdong Province (2018B030325002), Beijing Advanced Innovation Center for Future Chip (ICFC), and Tsinghua University Initiative Scientific Research Program.

\end{acknowledgments}

%



\end{document}